\pdfoutput=1
\documentclass[11pt,twoside,a4paper,cmspaper,final,collab]{cms-tdr}
\def\svnVersion{802bc29}\def\svnDate{2026/08/19}\def\cmsCernNoTag{CERN-EP-2025-169}\def\cmsCernDate{\today}\def\cmsMessage{Published in the Journal of High Energy Physics as \href{http://dx.doi.org/10.1007/JHEP08(2026)051}{\doi{10.1007/JHEP08(2026)051}.}}
\begin{document}\cmsNoteHeader{HIG-24-009}

\newcommand{\explimitonepoi}{506\xspace}
\newcommand{\explimittwopoi}{1097\xspace}
\newcommand{\expkappaconepoi}{95\xspace}
\newcommand{\obslimitonepoi}{1065\xspace}
\newcommand{\obslimittwopoi}{1804\xspace}
\newcommand{\obskappaconepoi}{211\xspace}
\newcommand{\obskappaccomb}{47\xspace}
\newcommand{\expkappaccomb}{51\xspace}
\newcommand{\obslimitcomb}{257}
\newcommand{\explimitcomb}{281}
\newcommand{\muonepoi}{\ensuremath{\mu_{\PQc\PH}=390^{+380}_{-270}}}
\newcommand{\mutwopoi}{\ensuremath{\mu_{\PQc\PH}=230^{+1000}_{-640}}}
\newcommand{\onepvalue}{1.5}
\newcommand{\kappac}{\ensuremath{\kappa_\PQc}\xspace}
\newcommand{\cH}{\ensuremath{\PQc\PH}\xspace}
\newcommand{\bH}{\ensuremath{\PQb\PH}\xspace}
\newcommand{\pp}{\ensuremath{\Pp\Pp}\xspace}
\newcommand{\ggH}{\ensuremath{\Pg\Pg\PH}\xspace}
\newcommand{\bdtbkg}{\ensuremath{\mathit{D}_\text{bkg}}\xspace}
\newcommand{\bdtH}{\ensuremath{\mathit{D}_\text{\PH{}-bkg}}\xspace}
\newcommand{\rch}{\ensuremath{\mu_{\PQc\PH}}\xspace}
\newcommand{\rbkghc}{\ensuremath{\mu_{\text{\PH{}+\PQc{}-bkg}}}\xspace}
\newcommand{\bkghc}{\ensuremath{\text{\PH{}+\PQc{}-bkg}}\xspace}
\newcommand{\bkgh}{\ensuremath{\text{\PH{}-bkg}}\xspace}
\newcommand{\sronej}{\ensuremath{N_\text{\PQc-j}=1}\xspace}
\newcommand{\srnj}{\ensuremath{N_\text{\PQc-j}>1}\xspace}
\newcommand{\mll}{\ensuremath{m_{\Pell\Pell}}\xspace}
\newcommand{\crmll}{\textit{high-\ensuremath{m_{\Pell\Pell}}}\xspace}
\providecommand{\cmsTable}[1]{\resizebox{\textwidth}{!}{#1}}
\newlength\cmsTabSkip\setlength{\cmsTabSkip}{1ex}
\providecommand{\cmsLeft}{left\xspace}
\providecommand{\cmsRight}{right\xspace}

\cmsNoteHeader{HIG-24-009}

\title{Search for a Higgs boson produced in association with a charm quark and  decaying to a \texorpdfstring{\PW}{W} boson pair in proton-proton collisions at \texorpdfstring{$\sqrt{s}=13\TeV$}{sqrt(s) = 13 TeV}}
       
\date{\today}

\abstract{
This paper presents a search for a Higgs boson produced in association with a charm quark ($\PQc\PH$) which allows to probe the Higgs-charm Yukawa coupling strength modifier \kappac. Higgs boson decays to a pair of \PW bosons are considered, where one \PW boson decays to an electron and a neutrino, and the other \PW boson decays to a muon and a neutrino. The data, corresponding to an integrated luminosity of 138\fbinv, were collected between 2016 and 2018 with the CMS detector at the LHC at a center-of-mass energy of $\sqrt{s}=13\TeV$. Upper limits at the 95\% confidence level (\CL) are set on the ratio of the measured yield to the standard model expectation for $\PQc\PH$ production. The observed (expected) upper limit is \obslimitonepoi\,(\explimitonepoi), corresponding to an observed (expected) constraint of $\abs{\kappac} < \obskappaconepoi\,(\expkappaconepoi)$. When combined with the previous search for $\PQc\PH$ in the diphoton decay channel of the Higgs boson, the limits are interpreted as observed (expected) constraints at 95\% \CL on the value of \kappac, $\abs{\kappac} < \obskappaccomb\,(\expkappaccomb)$. 
   }

\hypersetup{
pdfauthor={CMS Collaboration},
pdftitle={Search for a Higgs boson produced in association with a charm quark decaying to a W boson pair in proton-proton collisions at sqrt(s) = 13 TeV},
pdfsubject={CMS},
pdfkeywords={CMS, Higgs, rare production}}

\maketitle

\section{Introduction}\label{sec:intro}
Over a decade has passed since the ATLAS and CMS Collaborations at the CERN LHC discovered the Higgs (\PH) boson~\cite{:2012gu,:2012gk,CMS:2013btf}. With more data and improved analysis techniques, precise measurements of the properties of the \PH boson have become possible, along with its rare decay channels. The couplings of the \PH boson to other standard model (SM) particles are among its most important properties~\cite{CMS:2022dwd,ATLAS:2022vkf}. The coupling strength modifiers, $\kappa$, are introduced to parameterize potential deviations from the SM predictions in the \PH boson interactions with other particles. The couplings to bosons (\PZ, \PW, and \PGg) are well determined, based on precise measurements from the $\PH\to\PZ\PZ\to4\Pell$, $\PH\to\PW\PW$, and $\PH\to\PGg\PGg$ decay channels~\cite{CMS:2022dwd,ATLAS:2022vkf}. Among the Yukawa couplings, only those to third-generation fermions have been firmly established. The coupling modifiers $\kappa_\PQb$ and $\kappa_\PGt$ have been constrained through measurements of the $\PH\to\bbbar$~\cite{CMS:2023vzh,ATLAS:2020fcp,ATLAS:2024yzu} and $\PH\to\PGt\PGt$~\cite{ATLAS:2022yrq,CMS:2022kdi} processes, while $\kappa_\PQt$ has been constrained through a global fit~\cite{CMS:2022dwd,ATLAS:2022vkf}. The coupling to muons has been probed in the $\PH\to\PGm\PGm$ channel~\cite{CMS:2020xwi,ATLAS:2020fzp}, reaching evidence at the level of 3 standard deviations~\cite{CMS:2022dwd}. 

\begin{figure}[tbh]
    \centering
    \includegraphics[width=0.45\textwidth]{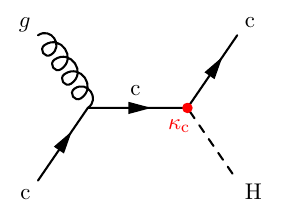}
    \includegraphics[width=0.45\textwidth]{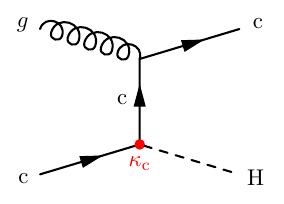}
    \caption{Leading-order Feynman diagrams that contribute to the $\pp\to\cH$ process. The red dots indicate vertices where the Higgs-charm coupling modifier \kappac is involved.}
    \label{fig:feynmann}
\end{figure}

The direct constraints on \kappac have been obtained through searches for $\PH\to\ccbar$ decays. 
These constraints have been derived in several production processes, including gluon fusion (\ggH)~\cite{CMS:2022fxs} and associated production with a vector boson $(\PV\PH)$~\cite{ATLAS:2022ers,CMS:2022psv,ATLAS:2024yzu}. 
With the advanced analysis techniques and recent improvements in charm jet identification methods (charm tagging) and calibrations~\cite{Bols:2020bkb,CMS:2021scf}, direct searches for $\PH\to \PQc\PQc$ decays have significantly improved in sensitivity. The most recent CMS result sets an observed (expected) constraint of $\abs{\kappac}<5.5$ (3.4) at 95\% confidence level (CL)~\cite{CMS:2022psv}. A recent ATLAS measurement further improves the observed (expected) limit to $\abs{\kappac}<4.2$ (4.1) at 95\% CL~\cite{ATLAS:2022ers,ATLAS:2024yzu}.  In addition to direct searches, \kappac can also be constrained indirectly through differential measurements of the transverse momentum (\pt) of the \PH boson across multiple decay channels~\cite{CMS:2025ihj,ATLAS:2022ers}.
An alternative method to probe the Higgs-charm coupling is via \PH boson production in association with charm quarks (\cH)~\cite{PhysRevD.100.073013,PhysRevLett.115.211801}. This channel probes \kappac via radiation of a \PH boson of a charm quark. Charm quarks appear abundant in the initial state together with a gluon. Leading-order (LO) Feynman diagrams contributing to the $\pp\to\cH$ process are shown in Fig.~\ref{fig:feynmann}. 
Despite a small cross section of this channel, corresponding to $\sim$0.2\% of the total \PH boson production rate, this channel offers the advantage for exploring simpler event signatures compared to those involving \PQc jets from the \PH boson decay. Recently, the CMS Collaboration reported on this production mechanism in the diphoton decay channel using the proton-proton (\pp) data set at $\sqrt{s}=13\TeV$ collected between 2016 and 2018~\cite{CMS:2025qmm}. An observed (expected) limit $\abs{\kappac}<38.1$ (72.5) at 95\% \CL has been obtained. Additionally, a search for the production of a \PH boson and one or more charm quarks was released by the ATLAS Collaboration~\cite{ATLAS:2024ext}. The ATLAS analysis did not measure the \cH component that is sensitive to \kappac. Hence, no constraint on $\abs{\kappac}$ was reported.

In this paper, a search for a \PH boson produced in association with a charm quark is presented, where the \PH boson decays to two \PW bosons, one decaying to an electron-neutrino pair and the other to a muon-neutrino pair. This final state is chosen to suppress background from {\PZ}+jets production, which has a significantly larger cross section and predominantly affects the same-flavor dilepton final states ($\Pe\Pe$\ and $\PGm\PGm$). Tabulated results are provided in the HEPData record for this analysis~\cite{hepdata}.

\section{The CMS detector}\label{sec:cms}

The central feature of the CMS apparatus is a superconducting solenoid of 6\unit{m} internal diameter, providing a magnetic field of 3.8\unit{T}. Within the solenoid volume are a silicon pixel and strip tracker, a lead tungstate crystal electromagnetic calorimeter (ECAL), and a brass and scintillator hadron calorimeter (HCAL), each composed of a barrel and two endcap sections. Forward calorimeters extend the pseudorapidity ($\eta$) coverage provided by the barrel and endcap detectors. Muons are reconstructed in gas-ionization detectors embedded in the steel flux-return yoke outside the solenoid. More detailed descriptions of the CMS detector, together with a definition of the coordinate system used and the relevant kinematic variables, can be found in Refs.~\cite{CMS:2008xjf,CMS:2023gfb}. 

Events of interest are selected using a two-tiered trigger system. The first level (L1), composed of custom hardware processors, uses information from the calorimeters and muon detectors to select events at a rate of around 100\unit{kHz} within a fixed latency of 4\mus~\cite{CMS:2020cmk}. The second level, known as the high-level trigger, consists of a farm of processors running a version of the full event reconstruction software optimized for fast processing and reduces the event rate to around 1\unit{kHz} before data storage~\cite{CMS:2016ngn,CMS:2024aqx}. During the 2016 and 2017 data-taking, a gradual shift in the timing of the inputs of the ECAL L1 trigger caused a specific trigger inefficiency. Correction factors were computed from data and applied to the simulated events, to mitigate this effect. 

\section{Data sets and simulated samples}\label{sec:sample}
The data sets used in the analysis were recorded by the CMS detector in the 2016, 2017, and 2018 data-taking periods, corresponding to an integrated luminosity of 36.3, 41.5, and 59.8\fbinv, respectively~\cite{CMS-LUM-17-003,CMS-PAS-LUM-17-004,CMS-PAS-LUM-18-002}. The 2016 data set is split into two separate periods, due to a substantial change in the tracker and muon system configurations in the middle of the data-taking, with the corresponding integrated luminosities of 19.5 and 16.8\fbinv, respectively. 
Events are selected by a combination of triggers requiring the presence of either one isolated muon with $\pt>24$ (27)\GeV for 2016 and 2018 (2017), or one isolated electron with \pt of 27, 35, and 32\GeV for 2016, 2017, and 2018, respectively, or the presence of both ($\Pe\PGm$), with values of $\pt>23$ (12)\GeV for the lepton with the highest (second-highest) \pt.

\begin{figure}
    \centering
    \includegraphics[width=0.40\linewidth]{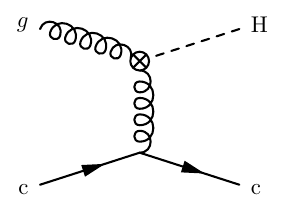}
    \caption{Diagram illustrating H boson production in association with a charm quark in the absence of charm Yukawa coupling. The crossed vertex represents the top quark loop.}
    \label{fig:bkgHc}
\end{figure}

Processes for \PH boson production in association with a \PQc or \PQb quark, \cH and \bH, are generated with \MGvATNLO~2.6~\cite{Alwall:2014hca,Frederix:2018nkq} at next-to-leading order (NLO).
Other \PH boson production processes, including \ggH, vector boson fusion (VBF), $\PV\PH$, and $\ttbar\PH$, are simulated with \POWHEG~v2~\cite{Nason:2004rx,Nason:2009ai,Frixione:2007vw,Alioli:2010xd,Bagnaschi:2011tu,Luisoni:2013cuh,Hartanto:2015uka}.
In this analysis, the \cH process refers to \PH boson production in association with at least one charm jet originating from diagrams proportional to the charm Yukawa coupling. Contributions from \PH boson production processes that yield charm jets but are not proportional to the charm Yukawa coupling, such as the one shown in Fig.~\ref{fig:bkgHc}, are treated as background.
Dedicated studies have been performed to assess the impact of heavy-quark masses in \cH and \bH processes.
In the 4-flavor scheme (4FS)~\cite{Wiesemann:2014ioa}, the calculations are performed with massless c quarks, while in the 3-flavor scheme (3FS) massive c quarks are assumed.  
 A similar approach is applied to \PQb quarks: in the 5-flavor scheme (5FS) massless b quarks are used, whereas in the 4FS their mass is taken into account. These two schemes are designed to model heavy quarks in different kinematic regimes, considering whether the mass of the quark plays a role of hard scattering scale in this process. The massless quark assumption simplifies the calculations significantly compared to including their masses, but this approach introduces divergences at low energies.
To mitigate this effect, dedicated events are generated in order to estimate the uncertainties arising from the kinematic differences.
This analysis assumes massless \PQc or \PQb quarks, using the 4FS for \cH and the 5FS for \bH. These calculations are combined with parton shower (PS) processes using FxFx jet merging~\cite{Frederix:2012ps} at NLO. The cross sections are found to be $\sigma_{\PQc\PH}=90\unit{fb}$ and $\sigma_{\PQb\PH}=660\unit{fb}$. The backward evolution recovers the information of the initial state \PQc and \PQb quarks~\cite{Wiesemann:2014ioa}.
The same method was used in our previous $\PQc\PH (\PH\to\gamma\gamma)$ search~\cite{CMS:2025qmm}.

All \PH boson production processes other than \cH are considered as background in this analysis and denoted as \bkgh.
The final state kinematics of the processes in Figs.~\ref{fig:feynmann} and \ref{fig:bkgHc} are very similar. Therefore, a multivariate discriminator based on a combination of kinematic variables is used to distinguish the processes represented by these diagrams, as discussed in Section~\ref{sec:strategy}. In addition, the modeling of \bkgh processes involving heavy-flavor jets poses significant theoretical challenges. To account for these difficulties, a 50\% uncertainty is assigned based on calculations for the {\ggH}+{\bbbar} process~\cite{Manzoni:2023qaf}.

Other background processes, such as top quark-antiquark (\ttbar), single top quark, and diboson ($\PW\PW,\PW\PZ,\PZ\PZ$) production are produced with \POWHEG~v2 at NLO. Production of \PZ bosons in association with jets ({\PZ}+jets) is simulated by \MGvATNLO~2.6 at NLO. For the \PW boson production in association with jets ({\PW}+jets), a combination of simulated events of LO and NLO from different generators (\MGvATNLO2.6 and \SHERPA2.2~\cite{Sherpa:2019gpd}) are used to obtain sufficiently large samples. The contributions of {\PW}+jets and {\PZ}+jets are denoted inclusively as {\PV}+jets in this paper.

Signal and background processes are simulated using various event generators. For all the simulated events, \PYTHIA~8.240~\cite{Sjostrand:2014zea} simulates underlying event (UE), PS, and hadronization. The simulations employ the NNPDF~3.1~\cite{NNPDF:2017mvq} parton distribution functions at next-to-next-to-leading order (NNLO), along with the CP5 underlying event tune~\cite{CMS:2019csb}.  The simulated events include additional \pp interactions within the same or nearby bunch crossings (pileup), which are corrected to match the distribution observed in data. The CMS detector response is modeled with \GEANTfour~\cite{Agostinelli:2002hh}.

\section{Event reconstruction and selection}\label{sec:reco}

The primary vertex is taken to be the vertex corresponding to the hardest scattering in the event, evaluated using tracking information alone, as described in Ref.~\cite{CMS-TDR-15-02}. Charged hadrons originating from other vertices (pileup vertices) are removed from the analysis. 

A particle-flow (PF) algorithm~\cite{CMS:2017yfk} aims to reconstruct and identify each individual particle in an event (PF candidate), with an optimized combination of information from the various elements of the CMS detector. 

Electrons are measured in $\abs{\eta} < 2.5$ with a gap at $1.44 < \abs{\eta} < 1.57$. The energy of electrons is determined from a combination of the track momentum at the main interaction vertex, the corresponding ECAL cluster energy, and the energy sum of all bremsstrahlung photons attached to the track. The electron momentum is estimated by combining the energy measurement in the ECAL with the momentum measurement in the tracker. The momentum resolution for electrons with $\pt\approx 45\GeV$ from $\PZ\to\Pe\Pe$ decays ranges from 1.6 to 5\%. It is generally better in the barrel region than in the endcaps, and also depends on the bremsstrahlung energy emitted by the electron as it traverses the material in front of the ECAL~\cite{CMS:2020uim,CMS-DP-2020-021}. Muons are measured in $\abs{\eta} < 2.4$, with detection planes made using three technologies: drift tubes, cathode strip chambers, and resistive-plate chambers. The momentum of muons is obtained from the corresponding track curvature. Matching muons to tracks measured in the silicon tracker results in a relative \pt resolution, for muons with \pt up to 100\GeV, of 1\% in the barrel and 3\% in the endcaps. The \pt resolution in the barrel is better than 7\% for muons with \pt up to 1\TeV~\cite{CMS:2018rym}. The selected electrons and muons are required to pass tight identification criteria, ensuring a low misidentification rate with lepton efficiencies of approximately 70\%~\cite{CMS:2018rym,CMS:2020uim}. To suppress leptons originating from \PQb and \PQc hadron decays, while retaining those from \PW boson decays, isolation from jet activity is required within a cone of radius $\Delta R =\sqrt{\smash[b]{(\Delta\eta)^2 + (\Delta\phi^2)}} = 0.3$ (0.4) for electrons (muons)~\cite{CMS:2018rym}, where $\phi$ is the azimuthal angle in radians. The isolation is defined as the scalar \pt sum of the PF candidates within the cone divided by the lepton \pt. 

Events are selected if they contain exactly one electron and one muon, with the leading lepton ($\ell 1$) $\pt > 25\GeV$ and subleading lepton ($\ell 2$) $\pt > 15\GeV$. A higher \pt threshold of 35 (30)\GeV is applied to the leading electron (muon) if an event is triggered only by a single-lepton trigger. Additionally, due to higher online \pt thresholds, electrons must have $\pt > 38\GeV$ in the 2017 data set.

For each event, hadronic jets are reconstructed by clustering PF candidates using the infrared- and collinear-safe anti-\kt algorithm~\cite{Cacciari:2008gp, Cacciari:2011ma} with a distance parameter of 0.4. Jet momentum is determined as the vectorial sum of all particle momenta in the jet, and is found from simulation to be, on average, within 5 to 10\% of the true momentum over the whole \pt spectrum and detector acceptance. Pileup can contribute additional tracks and calorimetric energy deposits to the jet momentum. To mitigate this effect, charged particles identified to be originating from pileup vertices are discarded and an offset correction is applied to correct for remaining contributions~\cite{CMS:2020ebo}. Jet energy corrections are derived from simulation to bring the measured response of jets to that of particle-level jets on average. In situ measurements of the momentum balance in dijet, photon+jet, {\PZ}+jet, and multijet events are used to account for any residual differences in the jet energy scale between data and simulation~\cite{CMS:2016lmd}. The jet energy resolution amounts typically to 15--20\% at 30\GeV, 10\% at 100\GeV, and 5\% at 1\TeV~\cite{CMS:2016lmd}. Additional selection criteria are applied to each jet to remove jets potentially dominated by anomalous contributions from various subdetector components or reconstruction failures.

Jets originating from charm quarks, \PQc jets, are identified using the \textsc{DeepJet} algorithm~\cite{Bols:2020bkb,CMS-DP-2018-058,CMS-DP-2023-006}. This is a deep neural network algorithm which utilizes several input features, including secondary-vertex properties, track-based variables, and PF jet constituents. Discrimination between \PQc jets and light jets (jets originating from \PQu, \PQd, and \PQs quarks or gluons) is achieved via the probability ratio defined as $CvsL = p(\PQc)/[p(\PQc) + p(\PQu\PQd\PQs\Pg)]$, where the $p(\PQc)$ and $p(\PQu\PQd\PQs\Pg)$ represent the classifier output for \PQc and light jets, respectively. Similarly, discrimination between \PQc jets and \PQb jets is based on the probability ratio defined as $CvsB = p(\PQc)/[p(\PQc) + p(\PQb)]$. These two discriminant ratios are the charm jet discriminants, which are used to select \PQc jet candidates with thresholds of $CvsL>0.12$ and $CvsB>0.5$. The threshold values are optimized for this analysis and provide a charm jet tagging efficiency of 43\% while the \PQb (light) jet mistag rate is 27\% (5\%). This allows discrimination between \cH and $\PQb\PH$ processes. The \PQc jet with the highest $CvsL$ score is considered the leading \PQc jet candidate. Events with at least one jet that passes the above charm jet selection criteria and that satisfies $\pt > 20\GeV$, $\abs{\eta} < 2.4$, and $\Delta R(\Pell,j) > 0.4$ (with $\Pell = \Pe, \PGm$) are selected. The shapes of \PQc jet discriminants in the simulated samples are corrected by flavor-dependent simulation-to-data efficiency scale factors~\cite{CMS:2021scf}. 

The missing transverse momentum (\ptvecmiss) is computed as the negative vector sum of the transverse momenta of all the PF candidates in an event, and its magnitude is denoted as \ptmiss~\cite{CMS:2019ctu}. The \ptvecmiss is modified to account for corrections to the energy scale of the reconstructed jets in the event. Only events passing $\ptmiss>20\GeV$ are selected. Anomalous high-\ptmiss events can arise due to a variety of reconstruction failures, detector malfunctions, or non-collision backgrounds. Such events are rejected by event filters that are designed to identify greater than 85\% of the spurious high-\ptmiss events with a mistagging rate less than 0.1\%~\cite{CMS:2019ctu}.

In this paper, only the $\PW\PW\to\Pe\PGn\PGm\PGn+\PQc$ final state is considered. The $\PW\PW\to\Pe\PGn\Pe\PGn$ and $\PGm\PGn\PGm\PGn$ decay channels are not considered due to the large {\PZ}+jet background contribution. Therefore, the presence of an electron and a muon with opposite charges and $\Delta R(\Pe,\PGm)>0.4$ is required. The invariant mass (the \pt) of the dilepton pair must be greater than 12 (30)\GeV to remove low mass resonances. Furthermore, the event selection criteria employ the transverse mass defined as
\begin{equation}
 \mT^{\text{obj}}=\sqrt{2 \pt^{\text{obj}}\ptmiss[1-\cos\Delta\phi(\text{obj},\ptvecmiss)]},
\end{equation}
where the $\pt^\text{obj}$ represents the transverse momentum of an object (a lepton, or dilepton system), and $\Delta\phi$ is the difference in $\phi$ between the object and \ptvecmiss. A requirement on the transverse mass with the subleading lepton, $\mT^{\Pell2} > 30\GeV$, is applied to suppress backgrounds not originating from \PW boson decays.

\section{Analysis strategy}\label{sec:strategy}

Events are categorized into two signal regions (SRs) and two control regions (CRs). The \PQc jet multiplicity requirement, \sronej and \srnj, defines two SRs with different sensitivities. In each SR, events must satisfy $\mT^{\Pell\Pell}>60\GeV$ and $\mll<72\GeV$ to maximize the signal efficiencies and to suppress the background contributions.
 Two CRs are defined, denoted as \crmll and \textit{top}. 
The \crmll CR follows the SR selection criteria, except for the dilepton invariant mass requirement, which is inverted to maintain a similar background composition as in the SRs. The \textit{top} CR, in addition to the inverted $\mll$ selection, requires at least two \PQc jet candidates and $\mT^{\Pell\Pell}< 60\GeV$ to enrich the contribution of \ttbar events. Table~\ref{tab:SRCR_summary} summarizes the definitions of the SRs and CRs. The dominant background process is \ttbar, which contributes approximately 90\% to the total background. 

\begin{table}[htpb]
\topcaption{Definitions of the SRs and CRs. Two SRs are defined according to \PQc jet multiplicity in selected events. Events with dilepton invariant mass above 72\GeV are considered as CRs. Events with an inverted transverse masses requirement that have at least two \PQc jets are considered in \textit{top} CR and the remaining events are tagged as \crmll CR.}
    \renewcommand{\arraystretch}{1.2}
   \cmsTable{
	\begin{tabular}{lccccc}
            Selection& &\sronej SR&\srnj SR& High $\mll$ CR&   Top CR\\\hline
        \PQc jet multiplicity &  $\text{N}_\text{\PQc-j}$ &$=$1  & $>$1  &$\geq$1 &$\geq$2   \\
        Dilepton invariant mass ({\GeVns}) & $m_{\ell\ell}$& $<$72 & $<$72 & $>$72& $>$72\\
        Transverse mass for the subleading lepton ({\GeVns}) &$\mT^{\ell2}$& $>$30 & $>$30 & $>$30& $>$30\\
        Transverse mass for the dilepton ({\GeVns})& $\mT^{\ell\ell}$& $>$60 & $>$60 & $>$60& $<$60\\
       	\end{tabular}
          }
    \label{tab:SRCR_summary}
\end{table}

\begin{figure}
    \centering
    \includegraphics[width=0.45\textwidth]{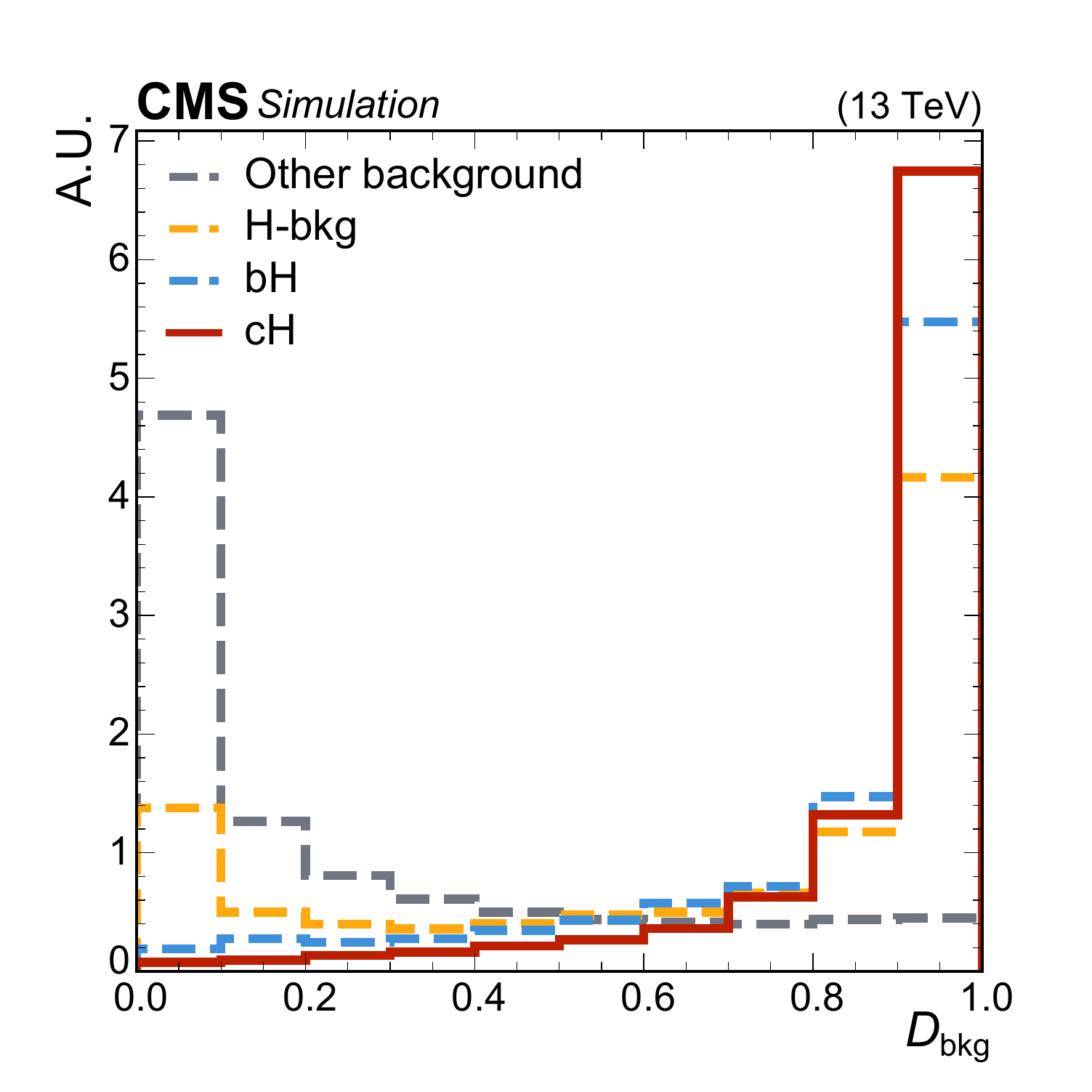}
    \includegraphics[width=0.45\textwidth]{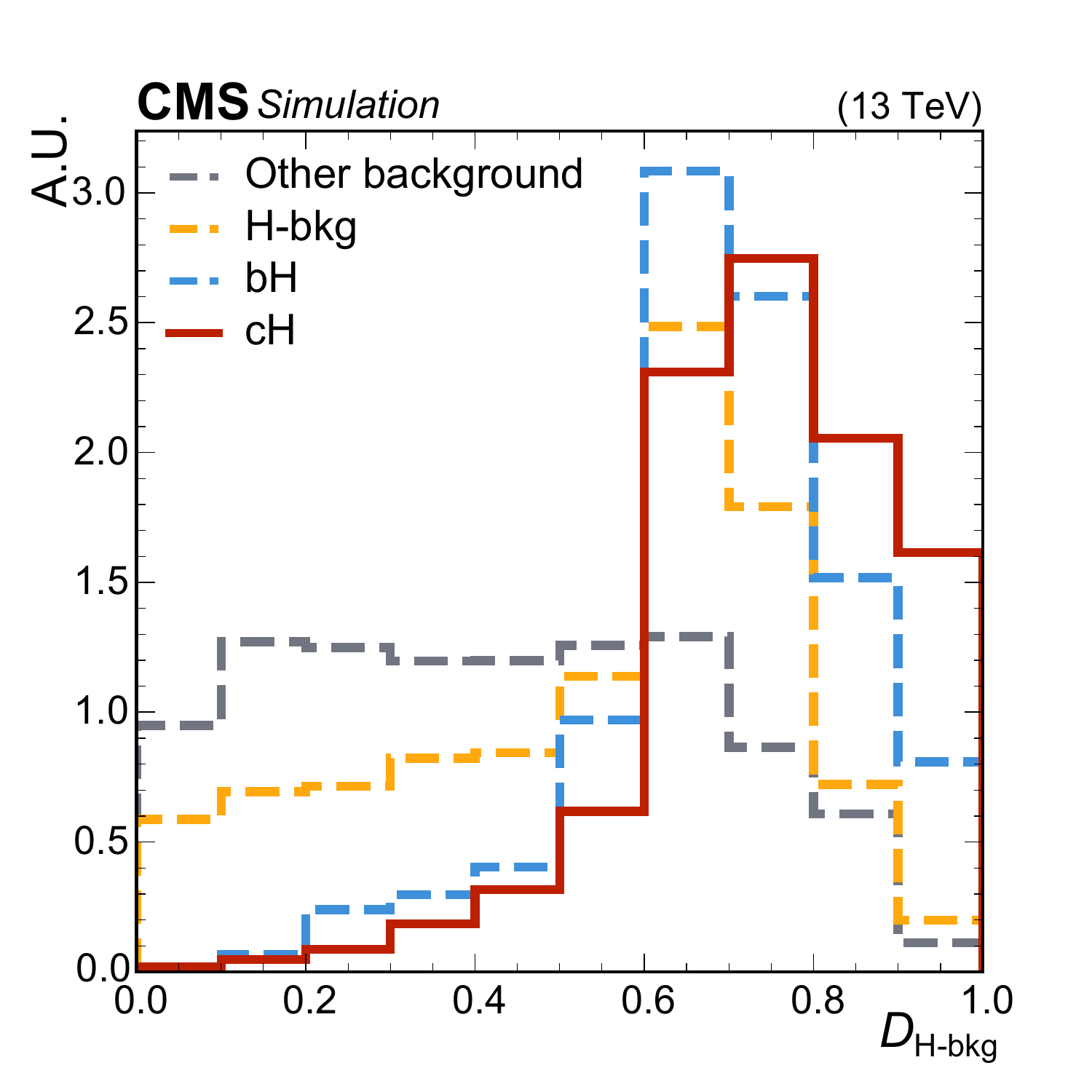}
    \caption{Area-normalized distributions of the two BDT classifiers: \bdtbkg (\cmsLeft) and \bdtH (\cmsRight). The H boson backgrounds (\bkgh) include \ggH, VBF, $\PV\PH$, $\ttbar\PH$, and \bH production. The non-{\PH}-boson backgrounds (\ttbar, single top, diboson, and {\PV}+jets) are collectively denoted as other background.}
    \label{fig:shape}
\end{figure}

A boosted decision tree (BDT) is used to construct two binary classifiers for distinguishing signal from background events. The BDT is trained inclusively across all signal regions. Owing to the limited statistical precision in the \srnj signal region, a dedicated training targeting this category was studied but found not to yield a significant improvement in sensitivity. The first classifier, \bdtbkg, is designed to disentangle the \cH process from the main background contributions (\ttbar, single top quark, diboson, and {\PV}+jets). The second classifier, \bdtH, is trained to separate the \cH signal from other \PH boson processes. Both BDTs use the same variables for training, including the \pt, the angular separations of the leptons, \ptmiss and the \PQc jet candidate, as well as the $CvsL$ and $CvsB$ values of the \PQc jet candidate. The full list of input variables to the BDT training are summarized in Table~\ref{tab:BDT_vars}.
The classifiers are implemented using the \textsc{XGBoost}~\cite{Chen:2016:XST:2939672.2939785} package with the gradient boosting method. The training data set includes 20\unit{k} events for the \cH signal, 40\unit{k} for \bkgh, and 100\unit{k} for the other background processes. These data sets are split equally between training and validation. Events are weighted according to their corresponding cross sections, and a class-balancing weight is applied to the signal to ensure balanced training. The BDT trainings are optimized separately for each data-taking era to account for differences in detector conditions. The performance of \bdtbkg and \bdtH is evaluated using the area under the ROC curve in each data-taking era, which ranges from 93 to 98\% for \bdtbkg and from 73 to 78\% for \bdtH. In \bdtbkg, kinematic and angular variables such as  $\Delta R(\Pell\Pell,\PQc)$ and $m_{\Pell\Pell}$ rank highest in importance, whereas in \bdtH the \textsc{DeepJet} $CvsL$ is among the most dominant inputs. The shapes of the two BDT classifiers are shown in Fig.~\ref{fig:shape}. The \bdtbkg classifier effectively separates the non-{\PH}-boson background contributions from the \cH production, but it is not sufficient for suppressing the \bkgh contributions. The dedicated BDT classifier, \bdtH, improves the \bkgh rejection by 12\% compared to \bdtbkg alone.

\begin{table}[!bth]
    \topcaption{Summary of input variables used in the BDT trainings.}
	\begin{tabular}{rl}
            Variable & Description \\\hline
            $\pt^{\Pell\Pell}$& Transverse momentum of dilepton system ($\Pe\PGm$)\\[\cmsTabSkip]
            $\pt^{\Pell1}$ & Transverse momentum of leading lepton (sorted by $\pt$) \\[\cmsTabSkip]
            $\pt^{\Pell2}$ & Transverse momentum of subleading lepton (sorted by $\pt$)\\[\cmsTabSkip]
            $\pt^{\PQc}$ & Transverse momentum of \PQc jet candidate\\[\cmsTabSkip]
            $\ptmiss$ & Missing transverse momentum\\[\cmsTabSkip]
             $\mll$& Mass of dilepton system ($\Pe\PGm$)\\[\cmsTabSkip]
            $\mT^{\Pell1}$ & Transverse mass of leading lepton and $\ptvecmiss$\\[\cmsTabSkip]
            $\mT^{\Pell2}$ & Transverse mass of subleading lepton and $\ptvecmiss$\\[\cmsTabSkip]
            $\Delta R(\Pell\Pell,\Pell1)$ & $\Delta R$ separation between dilepton system and leading lepton\\[\cmsTabSkip]
            $\Delta R(\Pell\Pell,\Pell2)$ & $\Delta R$ separation between dilepton system and subleading lepton\\[\cmsTabSkip]
            $\Delta R(\Pell\Pell,\PQc)$ & $\Delta R$ separation between dilepton system and \PQc jet\\[\cmsTabSkip]
            $\Delta\phi(\Pell \Pell+\ptvecmiss,\PQc)$ & $\Delta\phi$ separation of dilepton+\ptvecmiss system and \PQc jet\\[\cmsTabSkip]
            $\Delta\phi(\Pell 1,\ptvecmiss)$ & $\Delta\phi$ separation of leading lepton and \ptvecmiss\\[\cmsTabSkip]
            $\Delta\phi(\Pell 2,\ptvecmiss)$ & $\Delta\phi$ separation of subleading lepton and \ptvecmiss\\[\cmsTabSkip]
            \textsc{DeepJet} $CvsL$ & Discriminant separating light and \PQc jets\\[\cmsTabSkip]
            \textsc{DeepJet} $CvsB$ & Discriminant separating \PQb and \PQc jets\\[\cmsTabSkip]
            nSV & Number of secondary vertices in the event\\[\cmsTabSkip]
	\end{tabular}
\label{tab:BDT_vars}
\end{table}

\begin{figure}
    \centering
    \includegraphics[width=0.45\textwidth]{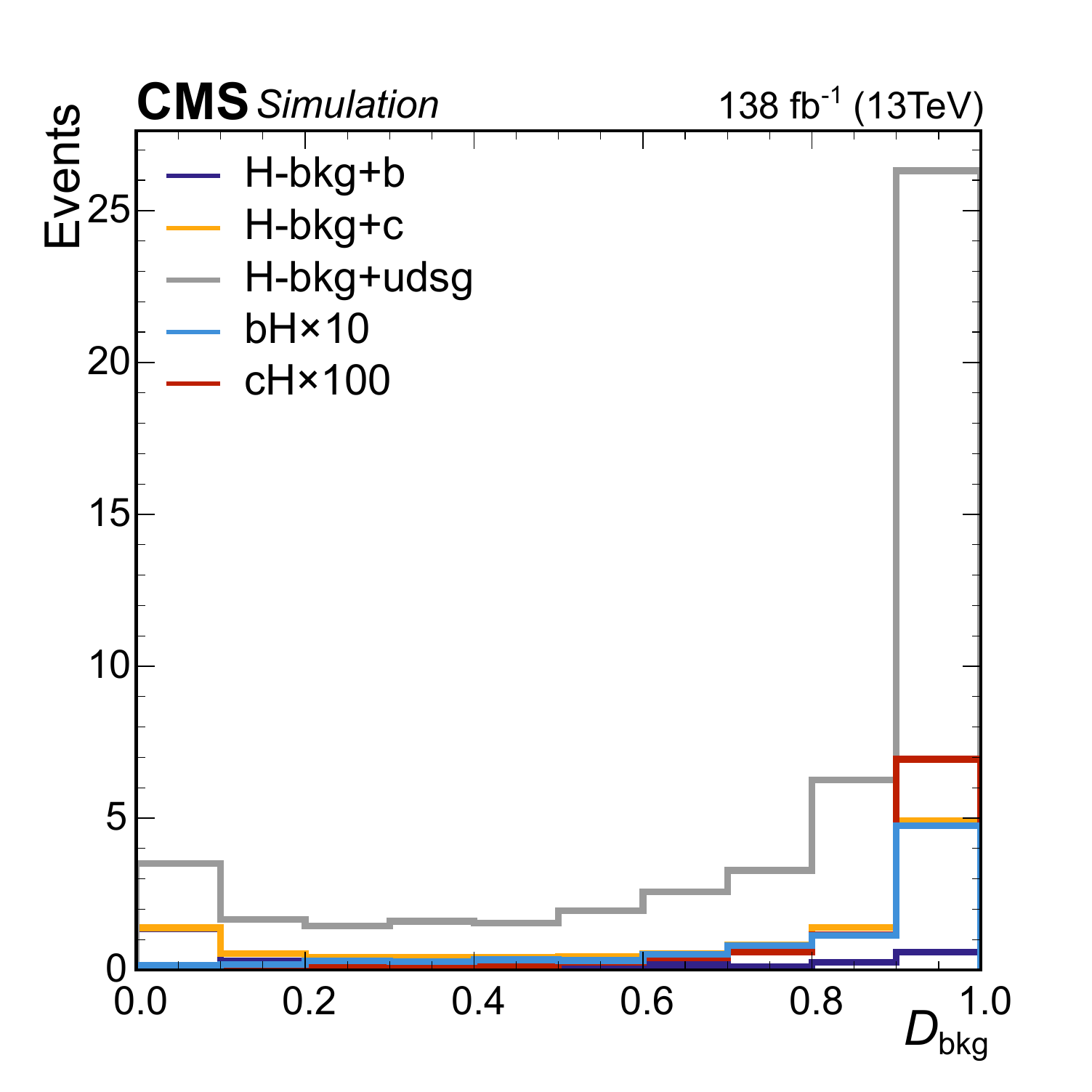}
    \includegraphics[width=0.45\textwidth]{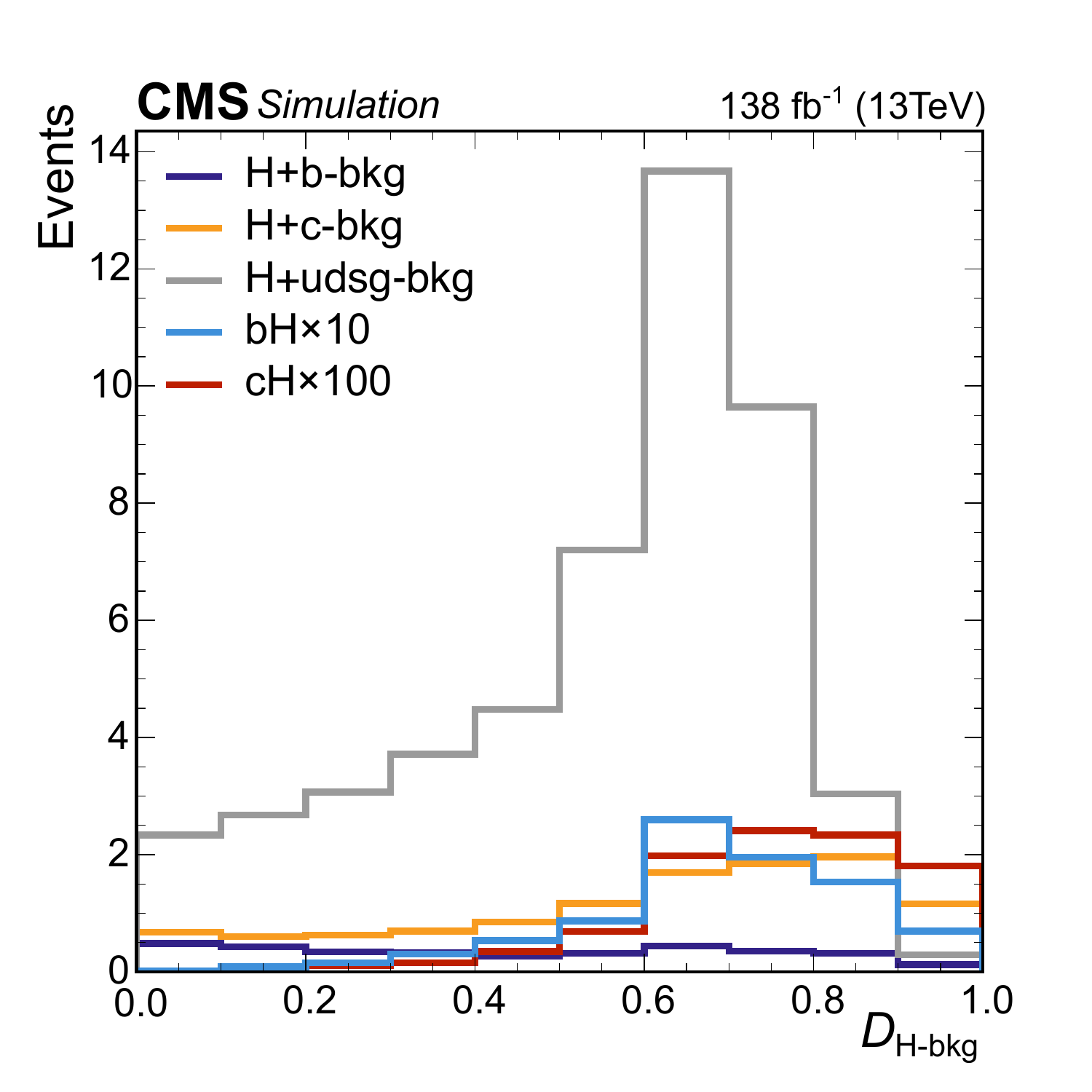}
    \caption{Distributions of \bdtbkg (\cmsLeft) and \bdtH (\cmsRight) for different \PH boson production processes. The event counts correspond to the expected yields for 138\fbinv. The splitting of \bkgh into three components is based on the flavor of the additional jet associated with the \PH boson. The yield of the \bH process is scaled by a factor of 10, and \cH  by a factor of 100.}
    \label{fig:Hc_split}
\end{figure}

Furthermore, \bkgh events are classified into three categories based on the generator-level flavor of the additional jet: \bkghc, {\PH}+\PQb-bkg, and {\PH}+$\PQu\PQd\PQs\Pg$-bkg, where the latter two terms are combined and denoted as {\PH}+not\PQc-bkg. The generator-level flavor of the jet is determined by ghost matching~\cite{Cacciari:2007fd}. Due to the presence of \PQc jets in the \bkghc component, its \bdtbkg distribution has a similar shape as the \cH process as shown in Fig.~\ref{fig:Hc_split} (\cmsLeft). However, its \bdtH shape is different.
The splitting of \bkgh into three components depending on the flavor of the associated jets is used in the signal extraction described later.

Two BDT classifiers, \bdtbkg and \bdtH, are combined to construct the distributions used for signal extraction. The \textit{k}-means~\cite{1056489,refId0} clustering algorithm is used to define the bin boundaries of the 2D BDT distribution, in order to preserve the information of the 2D BDT output while keeping the statistical uncertainties sufficiently small. 
The \textit{k}-means clustering algorithm partitions data into clusters with approximately equal variance by minimizing the distance between each point and its cluster centroid. 
Each resulting cluster is collapsed into a single bin, preserving its approximate location in the original 2D phase space. 
These bins are then reordered into a one-dimensional (1D) distribution, sorted by ascending BDT score according to their 2D position (\bdtH ascending first, \bdtbkg ascending second). The number of bins is optimized separately for each data-taking period and SR category to maximize sensitivity. For the \sronej category, 40 bins are used for the 2016 and 2017 data sets, and 50 bins for 2018 data set, due to its larger size. For the \srnj category, 30 bins are used across all years.

 The background contributions after event selection are estimated from both the CRs and SRs in data and simulated events. The normalizations for single top quark, diboson, and {\PV}+jets processes are obtained from simulated events, while the dominant background (\ttbar) is determined from data by a simultaneous fit in the SRs and CRs with unconstrained normalization factors. These normalization factors are determined separately for each data-taking period and \PQc jet multiplicity category. The \ttbar contribution in the \sronej SR is constrained by the normalization factor in the \crmll CR, whereas its contribution in \srnj SR is constrained from the \textit{top} CR due to the different charm jet multiplicity. In the fit, only the total event yield in the \textit{top} CR is used, while the binned \bdtH distribution is used in the \crmll CR. In the SRs, the clustered 1D distributions from \textit{k}-means binning are used. The \bH and {\PH}+not\PQc-bkg contributions are obtained from the simulation. Two scenarios are considered for the treatment of the \bkghc contribution. In the first scenario, the \bkghc contribution is set to the simulation expectation, within its uncertainties. In the second scenario, the \bkghc contribution is extracted simultaneously with $\PQc\PH$ signal, thus constrained from a fit to the data.

\section{Statistical procedure and systematic uncertainties}\label{sec:syst}

The statistical interpretation is performed using the CMS statistical analysis tool \textsc{Combine}~\cite{CAT-23-001}, which is based on the \textsc{RooFit}~\cite{Verkerke:2003ir} and \textsc{RooStats}~\cite{Moneta:2010pm} frameworks. Signal extraction is performed by a binned maximum likelihood fit to data in all SRs and CRs, using binned templates described in Section~\ref{sec:strategy}.

Systematic uncertainties affecting the results are categorized into theoretical and experimental. Furthermore, they are divided into two groups: uncertainties that only modify the normalization of the processes, and shape uncertainties that change the form of the fitted distributions. Some shape uncertainties, particularly in background templates, are poorly modeled due to the finite sample size of simulated events. To mitigate this, the relative deviations are smoothed using the \textsc{lowess} algorithm~\cite{lowess,doi:10.1080}, which preserves local structure. Smoothing is applied only to adjacent bins with correlated fluctuations in the 2D phase space.

The theoretical uncertainties are described in the following. For \cH and \bH processes, the FS uncertainty arises from differences in matrix-element calculations that treat the heavy quark in the proton PDF as either massive or massless. For \bH, the uncertainty is derived from the difference between the 5FS and 4FS; for \cH, it comes from the difference between 4FS and 3FS. These are treated as shape uncertainties and can reach up to 30\%, making them the dominant theoretical uncertainty in this analysis. A 50\% normalization uncertainty is assigned to the modeling of heavy-flavor contributions in other \PH boson production processes (\bkgh), based on the theoretical uncertainty of the {\ggH}+{\bbbar} process~\cite{Manzoni:2023qaf}. Production cross section uncertainties are treated as normalization uncertainties and are taken from theoretical predictions. A 1\% uncertainty is assigned to the single top quark production cross section~\cite{Campbell:2020fhf,PDF4LHCWorkingGroup:2022cjn,Kidonakis:2021vob}, 3\% for {\PV}+jets processes~\cite{Melnikov:2006kv,Catani:2009sm,Baglio:2022wzu,Anastasiou:2003ds,Dittmaier:2014qza}, 4\% for diboson production~\cite{Campbell:2010ff,Caola:2015rqy,Melia:2011tj,Campbell:2011bn,Cascioli:2014yka,Grazzini:2015hta,Kallweit:2018nyv,Lindert:2017olm}, and 1--3\% for \PH boson production and decay~\cite{deFlorian:2016spz}, depending on the production processes.
The UE and PS uncertainties, stemming from generator tuning and modeling choices, are included as shape uncertainties and typically amount to 3\%. The PDF uncertainties are evaluated by summing the 100 eigenvectors of the NNPDF set, contributing normalization uncertainties between 1 and 3\%, and up to 6\% for the \cH process. Renormalization and factorization scale uncertainties, resulting from independent variations of these two scales by factors of 0.5 and 2, are included as shape uncertainties with a typical size of 1\%. Lastly, \ttbar \pt mismodeling is incorporated as a shape uncertainty, based on reweighting the top quark \pt distribution to match NNLO predictions~\cite{Kidonakis:2012rm,Catani:2019hip,Czakon:2015owf,Czakon:2017wor}, with effects particularly significant in signal-enriched regions, reaching up to 3\%. The theoretical uncertainties are considered fully correlated across data-taking periods and have an overall impact of 66\% (46\%) on the 1POI (2POI) fit.

Experimental uncertainties are also assessed. Statistical uncertainties from limited simulated sample sizes are accounted for using the Barlow--Beeston-lite method~\cite{Conway:2011in}, assigning one parameter per bin across all processes, with typical magnitudes between 5 and 10\%. Jet energy scale and resolution uncertainties, propagated to jets and \ptmiss, are included as shape uncertainties and can reach up to 5\%, with enhancements in signal-enriched regions. Jet identification uncertainties are treated as shape variations, ranging between 1 and 2\% depending on the data-taking year. Charm jet identification uncertainties, derived from corrections to the \PQc jet discriminator distributions using control samples~\cite{CMS:2021scf}, are included as shape uncertainties and range from 3 to 7\%, with notable impact in signal-enriched regions. The integrated luminosity uncertainty ranges from 1.2 to 2.5\% across the 2016--2018 data-taking period~\cite{CMS-LUM-17-003,CMS-PAS-LUM-17-004,CMS-PAS-LUM-18-002}, with an overall uncertainty of 1.6\%, and is treated as a normalization uncertainty. Lepton trigger and identification efficiencies, including uncertainties from trigger, reconstruction, identification, and isolation corrections, are also treated as normalization uncertainties and are in the range between 1 and 3\%. A 1\% normalization uncertainty is assigned to account for L1 trigger inefficiency during 2016--2017. Pileup correction uncertainties are evaluated by varying the total inelastic cross section by 4.6\%~\cite{CMS:2018mlc}, and are treated as shape uncertainties. The statistical uncertainties of simulated events are treated as uncorrelated among all data-taking periods. Experimental uncertainties from \PQc jet identification, jet energy scale, jet energy resolution, and integrated luminosity are treated as partially correlated between data-taking years, while all other uncertainties are considered to be uncorrelated. The overall experimental uncertainties have a 46\% (48\%) impact on the 1POI (2POI) fit.

\section{Results}\label{sec:result}

\begin{figure}[htbp]
    \centering
    \includegraphics[width=0.45\textwidth]{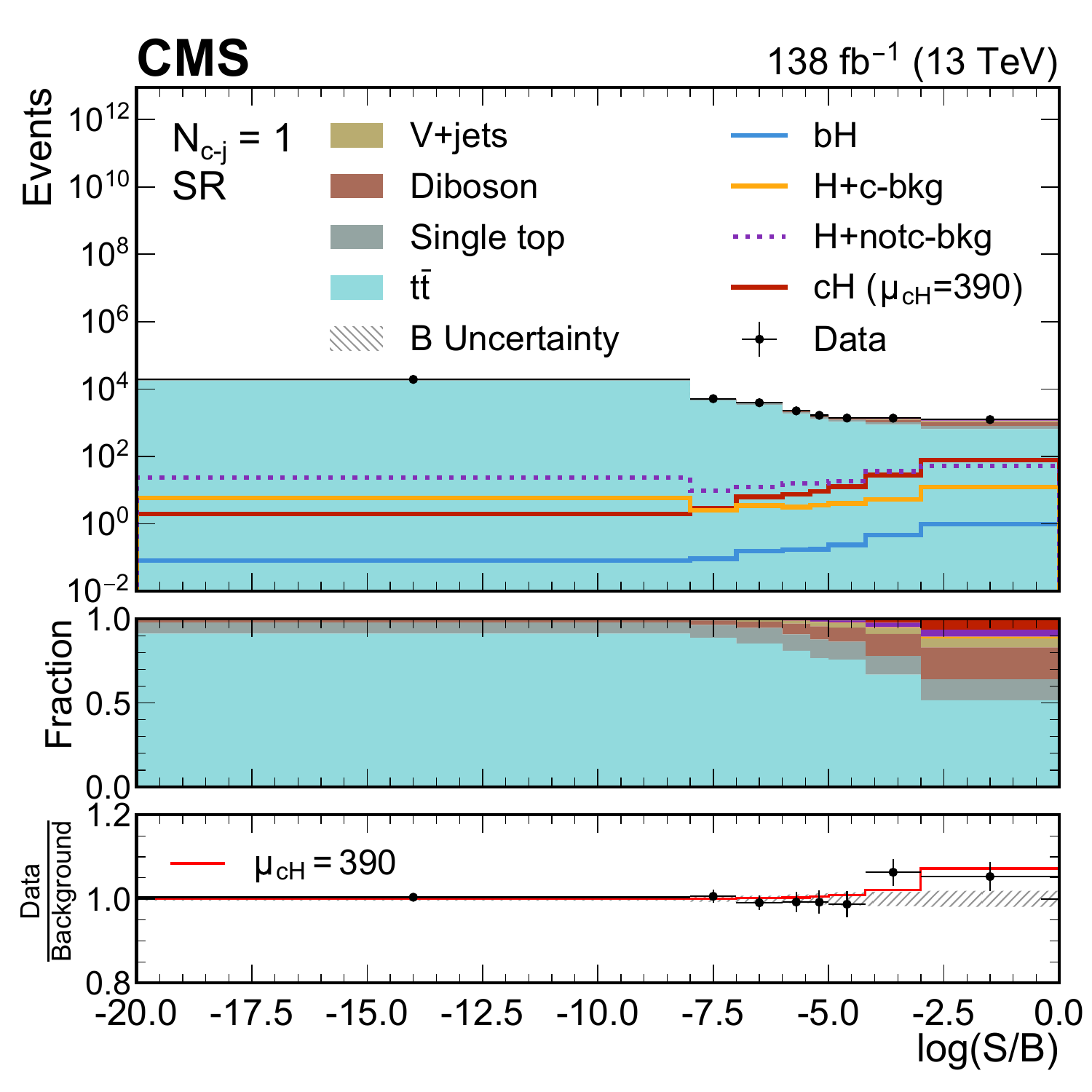}
    \includegraphics[width=0.45\textwidth]{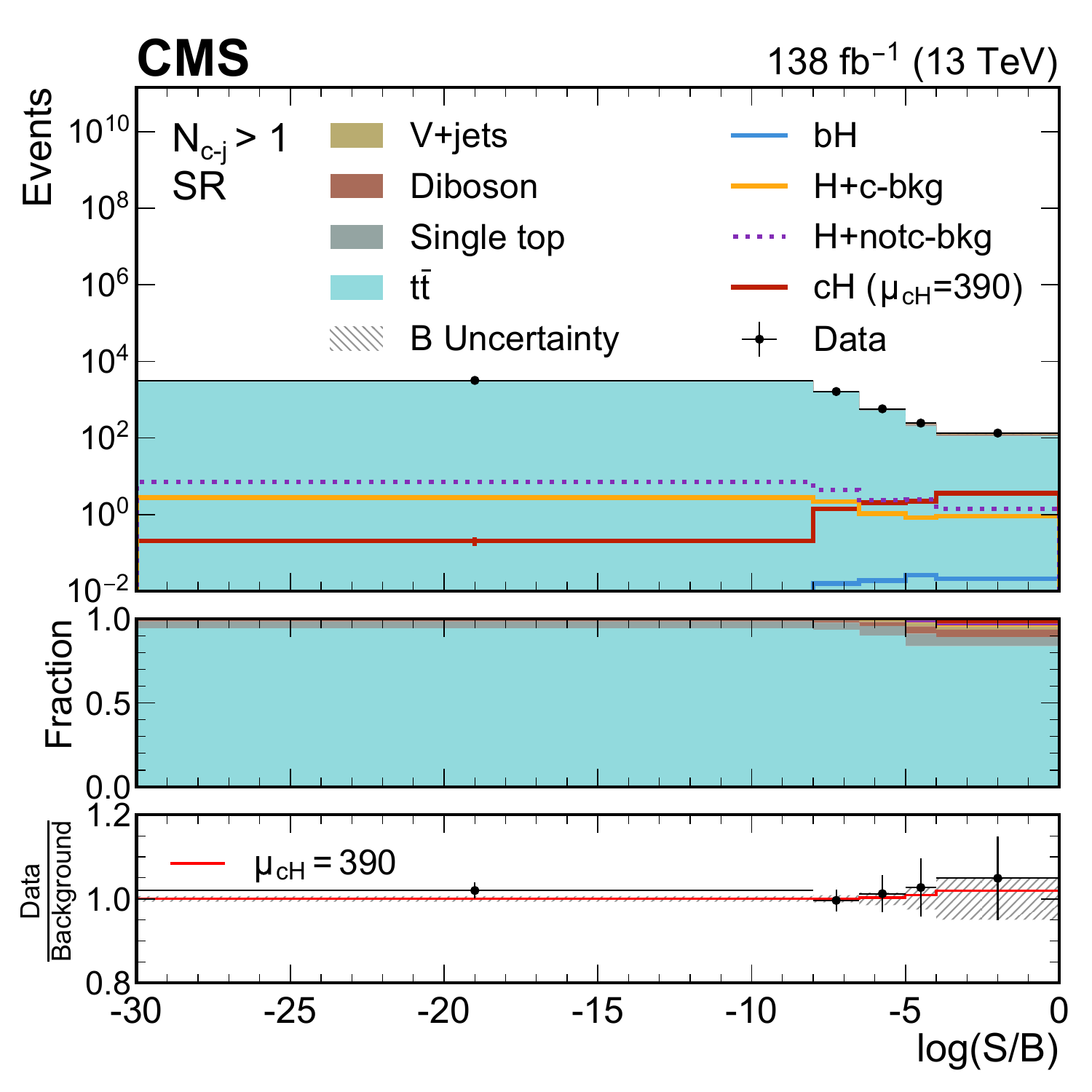}
    \includegraphics[width=0.45\textwidth]{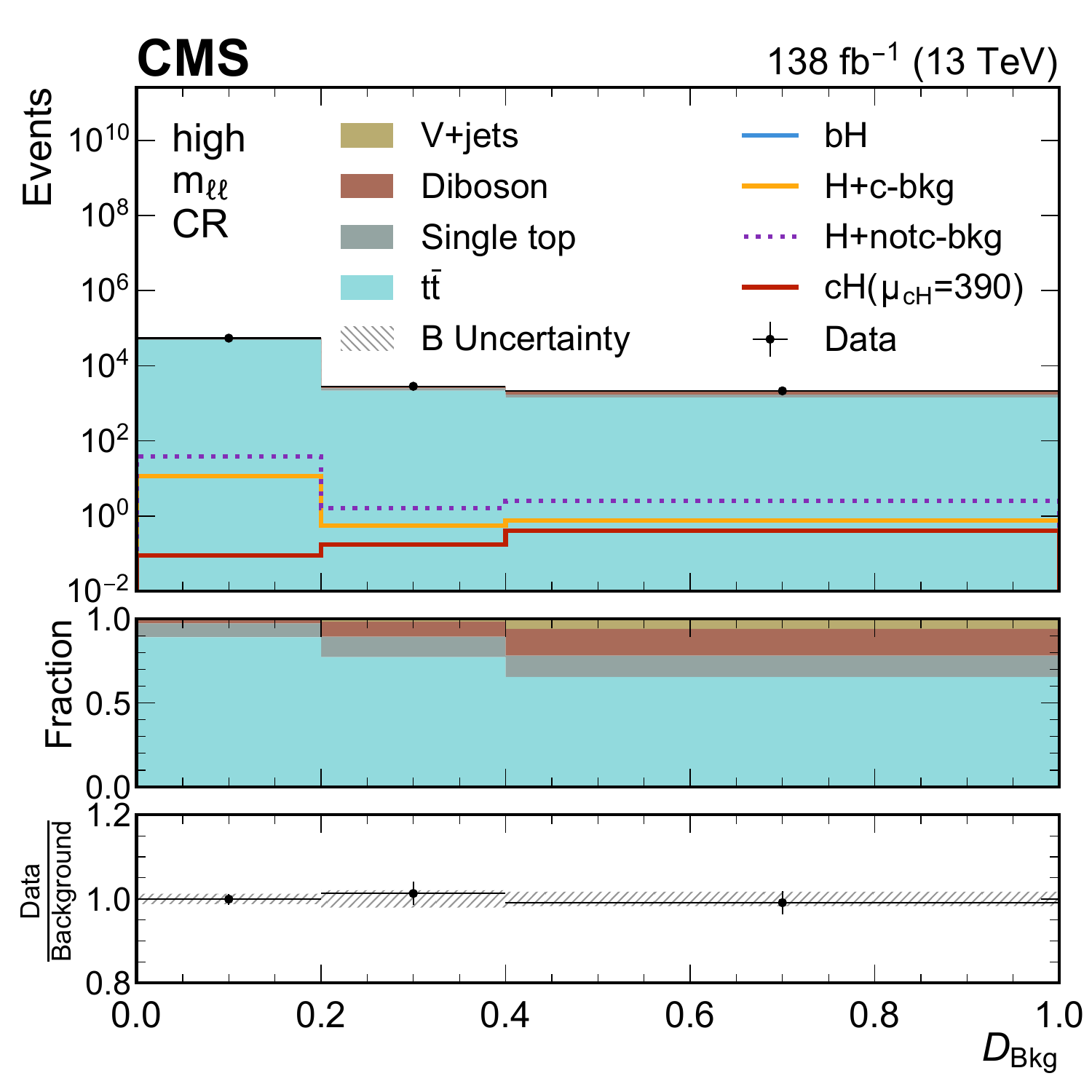}
    \includegraphics[width=0.45\textwidth]{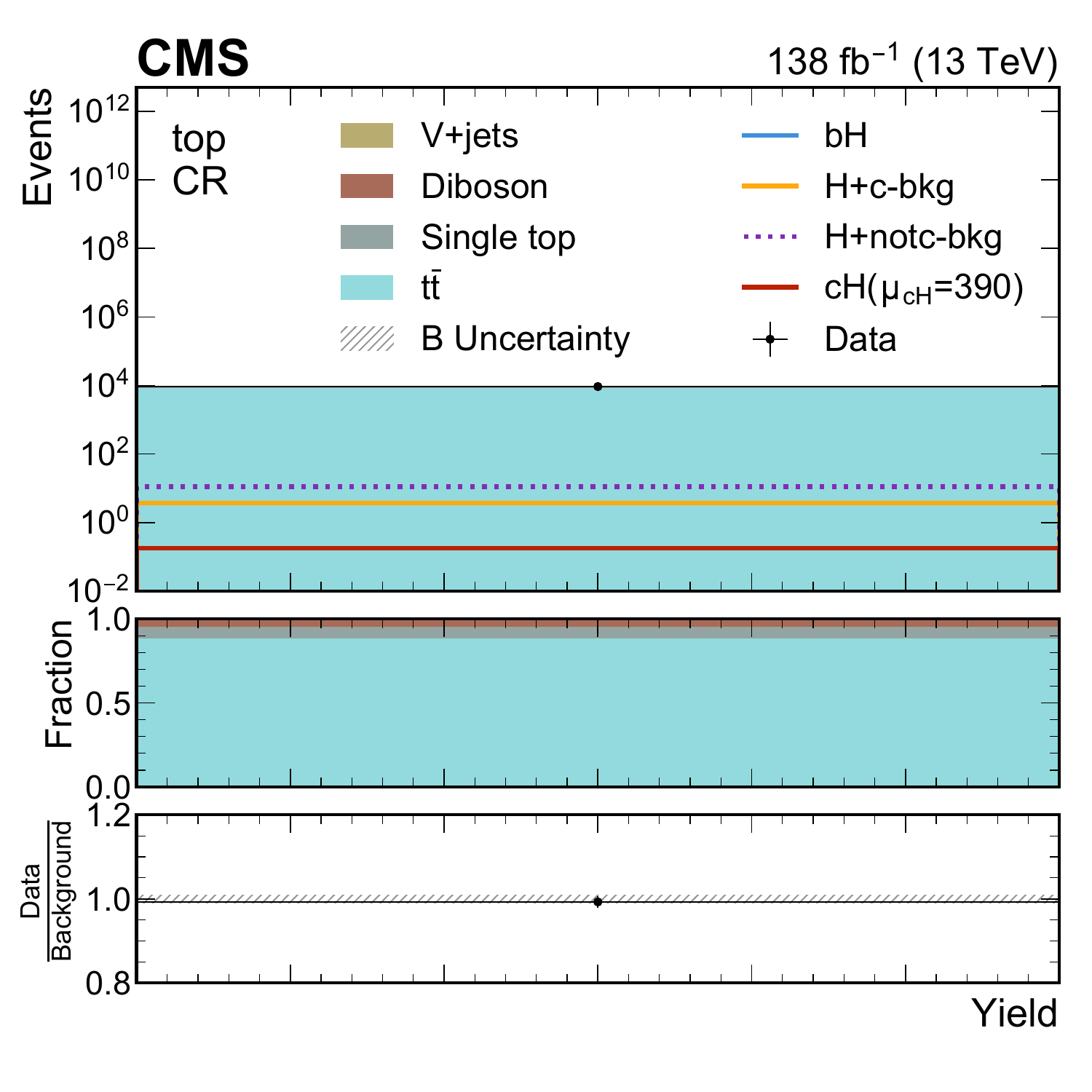}
    \caption{Observed and expected distributions of one-parameter-of-interest scenario after performing the fit in signal and control regions. 
    The upper left plot shows the distribution of events as a function of $\log(S/B)$ in the \sronej signal region, while the upper right shows the \srnj signal region, where $S$ and $B$ are the signal and background yields, respectively. 
    The lower left plot shows the distribution of \bdtbkg used in the \crmll control region, while the normalization factor of the \textit{top} CR is shown in the lower right. Within each plot, the upper panel provides the number of events from simulation and data in logarithmic scale, the middle panel displays the fractional contribution of each background component to the total expected yield in each bin, and the lower panel provides the ratio of data to simulated background events. The red lines represent background plus signal with $\rch=390$ divided by background. The hashed area shows the sum of systematic and statistical uncertainties of background events, which is denoted as B uncertainty.}
    \label{fig:BDT}
\end{figure}

The results are evaluated in two scenarios. In the first scenario, the contribution from \bkgh is fixed to its SM prediction. This corresponds to a one parameter of interest (1POI) fit, with the \cH signal strength modifier \rch which is defined as the ratio of measured yield to the one predicted by the SM. 
In the second scenario, the normalization of the charm jet associated \PH boson production, \rbkghc, is allowed to vary in the fit, resulting in a two parameter of interest (2POI) fit, with \rch and \rbkghc as parameters of interest. The fit is performed simultaneously in SRs and CRs using data collected between 2016 and 2018. Figure~\ref{fig:BDT} displays the observed and expected distributions of data and simulated events after performing the fit in the 1POI scenario.

\begin{figure}[htbp]
    \centering
    \includegraphics[width=0.7\textwidth]{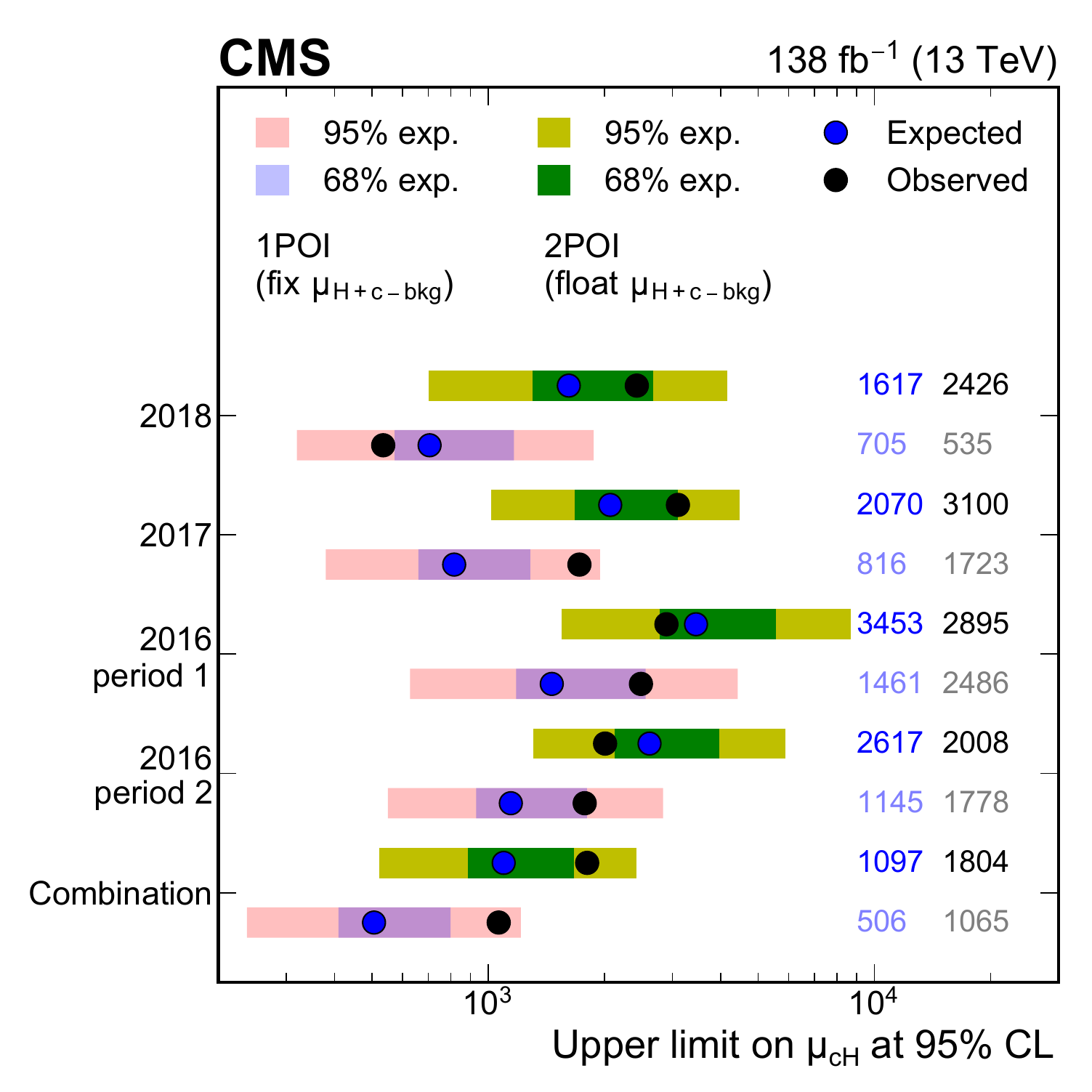}
    \caption{Upper limits of \rch at 95\% \CL for each data-taking period, and the combination of the periods. The light blue (red) bars show the 1 (2) standard deviation expected result of the 1POI fit with fixing the \bkghc contribution to the SM prediction. The green (yellow) bars show the 1 (2) standard deviation expected result of the 2POI fit with floating \bkghc contribution. The blue circles represent the median of the expected limit, while the black circles represent the observed limit.} 
    \label{fig:limit}
\end{figure}

\begin{figure}[htbp]
    \centering
    \includegraphics[width=0.7\textwidth]{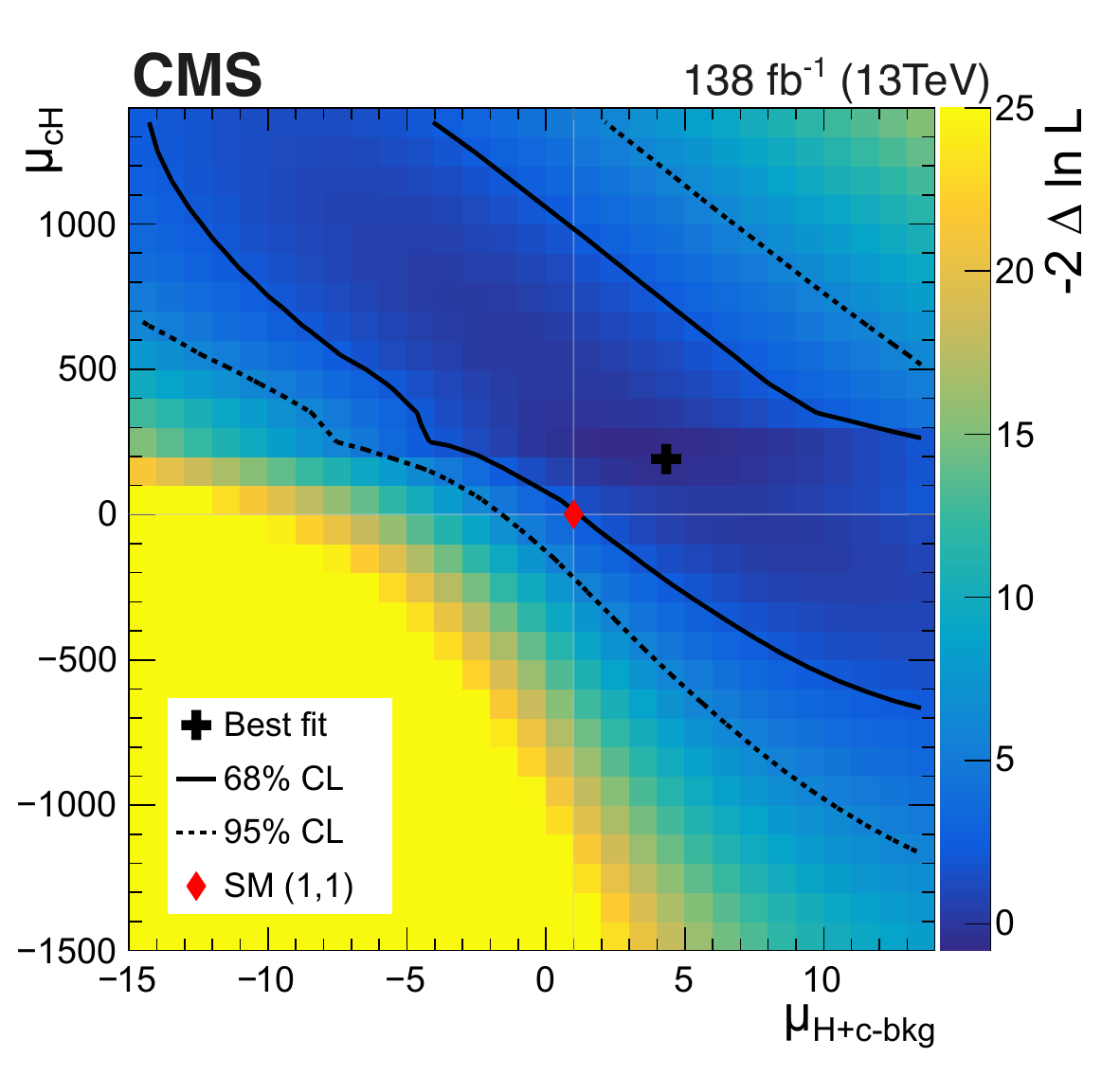}
    \caption{Two-dimensional likelihood contour of \rch and \rbkghc. The color scale represents twice the negative log likelihood difference with respect to the best fit point. The observed 95\% \CL(dashed) and 68\% \CL (solid) contours are shown in black lines, and the best fit point as a black cross. The SM expectation is marked by a red diamond. The kink in the contour arises from a local minimum in the likelihood, driven by the shape uncertainty associated with the flavor scheme used in the \cH process modeling.}
    \label{fig:2D}
\end{figure}

The observed (expected) upper limit on the \cH signal strength at 95\% \CL is \obslimitonepoi (\explimitonepoi) times the SM value in the 1POI fit. The difference between the observed and expected limits is primarily driven by an upward fluctuation in the 2017 data in the \sronej category, which leads to a slightly weaker observed limit. The sensitivity is dominated by the \sronej category, which is approximately four times more sensitive than the \srnj category. In the 2POI fit, the corresponding upper limit is \obslimittwopoi (\explimittwopoi) times the SM prediction.
The measured signal strength for the \cH process in the 1POI fit is found to be \muonepoi, which is compatible with the SM within \onepvalue~standard deviations. The results from different data-taking periods are summarized in Fig.~\ref{fig:limit}. 
A strong correlation of $-87\%$ is observed between \bkghc and \cH productions in the 2POI fit, leading to a weaker constraint of \mutwopoi, as shown in Fig.~\ref{fig:2D}. Similar kinematic properties of $\PQb\PH$ and \cH production also result in a strong correlation of $-89\%$ in the fit.

The sensitivity of this analysis is primarily limited by statistical uncertainties in data, with nonnegligible contributions from theoretical uncertainties, such as the FS choice of $\PQc\PH$ production and heavy-flavor modeling in \ggH production. The impact of each uncertainty on the measurement of \rch is summarized in Table~\ref{tab:split_unc}.

The resulting upper limit on \rch is translated into a constraint on \kappac, using the flat direction assumption~\cite{PhysRevLett.115.211801,Aguilar-Saavedra:2020rgo}. Under this assumption, the modifier of all other \PH couplings $\kappa_\PH$ changes according to \kappac, such as to keep the signal strengths of other \PH boson processes at unity. The observed (expected) interval at 95\% \CL is constrained to $\abs{\kappac}<\obskappaconepoi$ (\expkappaconepoi).

\begin{figure}[htbp]
    \centering
    \includegraphics[width=0.7\textwidth]{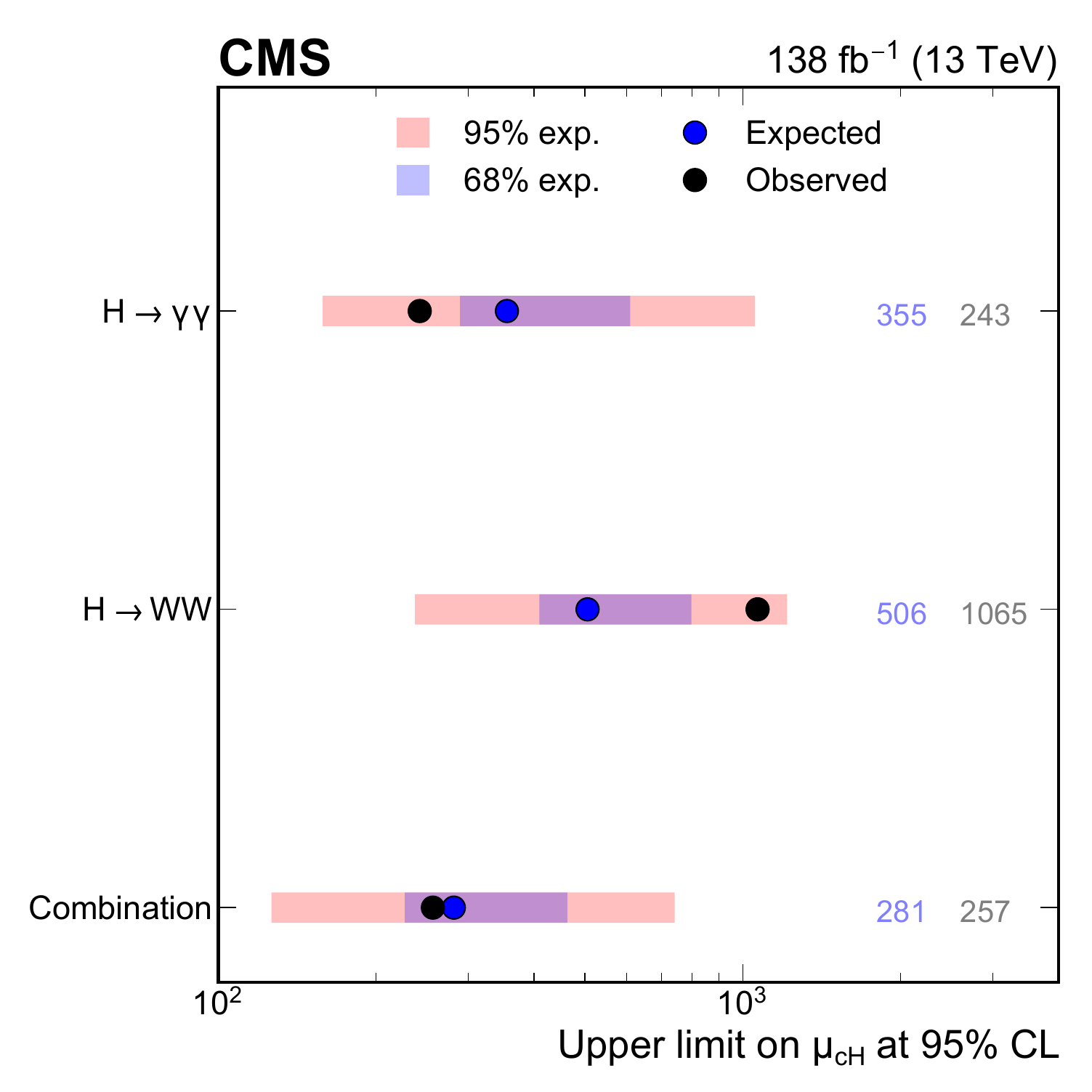}
    \caption{Upper limits of \rch at 95\% \CL of the combined analysis with the previous result in the diphoton channel~\cite{CMS:2025qmm} using the 1POI fit. The light blue (red) bars show the 1 (2) standard deviation range of the expected results. The blue circles represent the median of the expected limit, while the black circles represent the observed limit.} 
    \label{fig:combine}
\end{figure}

A combined analysis of the 1POI fit with the previous search in the diphoton decay channel~\cite{CMS:2025qmm} is performed. Common experimental and theoretical uncertainties related to \PH boson production cross sections are treated as correlated. The combination improves the expected upper limit on \rch at 95\% \CL to \explimitcomb, while the observed limit is \obslimitcomb, as shown in Fig.~\ref{fig:combine}. This leads to constraints on the observed (expected) value of $\abs{\kappac}<\obskappaccomb$ (\expkappaccomb) at 95\% \CL. 

Looking ahead to the High-Luminosity LHC, larger datasets will substantially reduce statistical uncertainties. In parallel, advances in heavy-flavor jet identification, including modern transformer-based algorithms~\cite{CMS-DP-2024-066}, already show improvements in light and b jet rejection of up to a factor of two. Continued progress in the theoretical modeling of heavy-flavor production will also be essential to maximize sensitivity.

\begin{table}[htbp]
    \centering
    \topcaption{Relative impact (in percent) of individual uncertainty sources on the measurement of \rch expressed as the change in the one standard deviation when each source is fixed. Statistical uncertainties are derived by fixing all constrained nuisance parameters.}
    \begin{tabular}{lcc}
        Uncertainty source   &  1POI  &2POI \\\hline        
        \multicolumn{3}{c}{Theoretical}\\[\cmsTabSkip]
        \cH modeling &   22.1 & 10.9 \\
        \PH-bkg  &21.4 & 10.0 \\
        Other backgrounds &  4.5 & 3.5 \\[\cmsTabSkip]
        \multicolumn{3}{c}{Experimental}\\[\cmsTabSkip]
        MC statistical (bin by bin)   & 6.2 & 9.1 \\
        Pileup correction & 4.3 & 1.2\\
        Lepton efficiencies & 4.6 & 3.2  \\
        Jet energy scale and resolution & 6.0 & 7.7 \\
        Charm jet identification	&  5.9 & 4.1\\
      \ttbar normalization	&  5.6 & 2.0\\[\cmsTabSkip]
        Statistical uncertainty  &  60.0 & 77.1 \\
      \end{tabular}
    \label{tab:split_unc}
\end{table}

\clearpage
\section{Summary}
The first search for the associated production of a charm quark and a Higgs boson (\cH) with the Higgs boson decays to a pair of \PW bosons has been presented. Decays of the \PW boson pair to $\Pe\PGn\PGm\PGn$ are considered. The search is based on the data collected from 2016 to 2018 with the CMS detector at the LHC, corresponding to an integrated luminosity of 138\fbinv. The observed (expected) upper limit at 95\% confidence level (\CL) on the ratio of the measured production cross section with respect to the value expected from the standard model (SM) for \cH production \rch is set at \obslimitonepoi (\explimitonepoi) times the SM prediction. This search provides an alternative probe of the Yukawa coupling between the Higgs boson and charm quark. When combined with the previous search for \cH in the diphoton decay channel, the limits are interpreted as observed (expected) constraints on the Yukawa coupling modifier of the Higgs boson to the charm quark, yielding $\abs{\kappac} < \obskappaccomb$ (\expkappaccomb) at 95\% \CL.

\begin{acknowledgments}
We congratulate our colleagues in the CERN accelerator departments for the excellent performance of the LHC and thank the technical and administrative staffs at CERN and at other CMS institutes for their contributions to the success of the CMS effort. In addition, we gratefully acknowledge the computing centers and personnel of the Worldwide LHC Computing Grid and other centers for delivering so effectively the computing infrastructure essential to our analyses. Finally, we acknowledge the enduring support for the construction and operation of the LHC, the CMS detector, and the supporting computing infrastructure provided by the following funding agencies: SC (Armenia), BMBWF and FWF (Austria); FNRS and FWO (Belgium); CNPq, CAPES, FAPERJ, FAPERGS, and FAPESP (Brazil); MES and BNSF (Bulgaria); CERN; CAS, MoST, and NSFC (China); MINCIENCIAS (Colombia); MSES and CSF (Croatia); RIF (Cyprus); SENESCYT (Ecuador); ERC PRG, TARISTU24-TK10 and MoER TK202 (Estonia); Academy of Finland, MEC, and HIP (Finland); CEA and CNRS/IN2P3 (France); SRNSF (Georgia); BMFTR, DFG, and HGF (Germany); GSRI (Greece); NKFIH (Hungary); DAE and DST (India); IPM (Iran); SFI (Ireland); INFN (Italy); MSIT and NRF (Republic of Korea); MES (Latvia); LMTLT (Lithuania); MOE and UM (Malaysia); BUAP, CINVESTAV, CONACYT, LNS, SEP, and UASLP-FAI (Mexico); MOS (Montenegro); MBIE (New Zealand); PAEC (Pakistan); MES, NSC, and NAWA (Poland); FCT (Portugal);  MESTD (Serbia); MICIU/AEI and PCTI (Spain); MOSTR (Sri Lanka); Swiss Funding Agencies (Switzerland); MST (Taipei); MHESI and NSTDA (Thailand); TUBITAK and TENMAK (T\"{u}rkiye); NASU (Ukraine); STFC (United Kingdom); DOE and NSF (USA).

\hyphenation{Rachada-pisek} Individuals have received support from the Marie-Curie program and the European Research Council and Horizon 2020 Grant, contract Nos.\ 675440, 724704, 752730, 758316, 765710, 824093, 101115353, 101002207, 101001205, and COST Action CA16108 (European Union); the Leventis Foundation; the Alfred P.\ Sloan Foundation; the Alexander von Humboldt Foundation; the Science Committee, project no. 22rl-037 (Armenia); the Fonds pour la Formation \`a la Recherche dans l'Industrie et dans l'Agriculture (FRIA-Belgium); the Beijing Municipal Science \& Technology Commission, No. Z191100007219010, the Fundamental Research Funds for the Central Universities, the Ministry of Science and Technology of China under Grant No. 2023YFA1605804, and the Natural Science Foundation of China under Grant No. 12061141002 (China); the Ministry of Education, Youth and Sports (MEYS) of the Czech Republic; the Shota Rustaveli National Science Foundation, grant FR-22-985 (Georgia); the Deutsche Forschungsgemeinschaft (DFG), among others, under Germany's Excellence Strategy -- EXC 2121 ``Quantum Universe" -- 390833306, and under project number 400140256 - GRK2497; the Hellenic Foundation for Research and Innovation (HFRI), Project Number 2288 (Greece); the Hungarian Academy of Sciences, the New National Excellence Program - \'UNKP, the NKFIH research grants K 131991, K 133046, K 138136, K 143460, K 143477, K 146913, K 146914, K 147048, 2020-2.2.1-ED-2021-00181, TKP2021-NKTA-64, and 2021-4.1.2-NEMZ\_KI-2024-00036 (Hungary); the Council of Science and Industrial Research, India; ICSC -- National Research Center for High Performance Computing, Big Data and Quantum Computing, FAIR -- Future Artificial Intelligence Research, and CUP I53D23001070006 (Mission 4 Component 1), funded by the NextGenerationEU program (Italy); the Latvian Council of Science; the Ministry of Education and Science, project no. 2022/WK/14, and the National Science Center, contracts Opus 2021/41/B/ST2/01369, 2021/43/B/ST2/01552, 2023/49/B/ST2/03273, and the NAWA contract BPN/PPO/2021/1/00011 (Poland); the Funda\c{c}\~ao para a Ci\^encia e a Tecnologia, grant CEECIND/01334/2018 (Portugal); the National Priorities Research Program by Qatar National Research Fund; MICIU/AEI/10.13039/501100011033, ERDF/EU, "European Union NextGenerationEU/PRTR", and Programa Severo Ochoa del Principado de Asturias (Spain); the Chulalongkorn Academic into Its 2nd Century Project Advancement Project, and the National Science, Research and Innovation Fund via the Program Management Unit for Human Resources \& Institutional Development, Research and Innovation, grant B39G680009 (Thailand); the Kavli Foundation; the Nvidia Corporation; the SuperMicro Corporation; the Welch Foundation, contract C-1845; and the Weston Havens Foundation (USA).
\end{acknowledgments}
\bibliography{auto_generated} 

\cleardoublepage \appendix\section{The CMS Collaboration \label{app:collab}}\begin{sloppypar}\hyphenpenalty=5000\widowpenalty=500\clubpenalty=5000
\cmsinstitute{Yerevan Physics Institute, Yerevan, Armenia}
{\tolerance=6000
A.~Hayrapetyan, V.~Makarenko\cmsorcid{0000-0002-8406-8605}, A.~Tumasyan\cmsAuthorMark{1}\cmsorcid{0009-0000-0684-6742}
\par}
\cmsinstitute{Institut f\"{u}r Hochenergiephysik, Vienna, Austria}
{\tolerance=6000
W.~Adam\cmsorcid{0000-0001-9099-4341}, J.W.~Andrejkovic, L.~Benato\cmsorcid{0000-0001-5135-7489}, T.~Bergauer\cmsorcid{0000-0002-5786-0293}, M.~Dragicevic\cmsorcid{0000-0003-1967-6783}, C.~Giordano, P.S.~Hussain\cmsorcid{0000-0002-4825-5278}, M.~Jeitler\cmsAuthorMark{2}\cmsorcid{0000-0002-5141-9560}, N.~Krammer\cmsorcid{0000-0002-0548-0985}, A.~Li\cmsorcid{0000-0002-4547-116X}, D.~Liko\cmsorcid{0000-0002-3380-473X}, M.~Matthewman, I.~Mikulec\cmsorcid{0000-0003-0385-2746}, J.~Schieck\cmsAuthorMark{2}\cmsorcid{0000-0002-1058-8093}, R.~Sch\"{o}fbeck\cmsAuthorMark{2}\cmsorcid{0000-0002-2332-8784}, D.~Schwarz\cmsorcid{0000-0002-3821-7331}, M.~Shooshtari, M.~Sonawane\cmsorcid{0000-0003-0510-7010}, W.~Waltenberger\cmsorcid{0000-0002-6215-7228}, C.-E.~Wulz\cmsAuthorMark{2}\cmsorcid{0000-0001-9226-5812}
% ...
\par}
\vskip\cmsinstskip
\dag:~Deceased\\
$^{1}$Also at Yerevan State University, Yerevan, Armenia\\
$^{2}$Also at TU Wien, Vienna, Austria\\
% ...
\end{sloppypar}
%%% END EDITABLE REGION %%%
% skeleton_end
\end{document}